\documentclass[12pt]{iopart}
\pdfoutput=1

\usepackage{url}
\usepackage{iopams}
\usepackage{hyperref}
\usepackage{amssymb}
\usepackage{bm}
\usepackage{graphicx}
\usepackage{xcolor}
\usepackage{setspace}
\usepackage{graphicx}
\usepackage{subfigure} 
\usepackage{algorithm}
\usepackage{algorithmic}
\usepackage{localmacros}
\usepackage{wrapfig}
\usepackage{xspace}

\usepackage{tikz}
\usetikzlibrary{shapes,arrows,decorations.pathreplacing,positioning}
\tikzset{%
	factor/.style = {draw, thick, rectangle, minimum height = 1em, minimum width = 1em, fill = black},
	var/.style = {draw, thick, circle, minimum height = 2em, minimum width = 2em},
	link/.style = {coordinate}
}
\newcommand{\tikzImageOneX}{0}
\newcommand{\tikzImageTwoX}{0.75}
\newcommand{\tikzImageThreeX}{1.5}
\newcommand{\tikzImageFourX}{2.25}
\newcommand{\tikzImageFiveX}{3}
\newcommand{\tikzImageSixX}{3.75}
\newcommand{\tikzImageSevenX}{4.5}
\newcommand{\tikzImageEightX}{5.25}
\newcommand{\tikzImageNineX}{6.0}
\newcommand{\tikzImageTenX}{6.75}
\newcommand{\tikzImageRowOneY}{0}
\newcommand{\tikzImageRowTwoY}{-0.8}
\newcommand{\tikzImageRowThreeY}{-1.6}
\newcommand{\tikzImageRowFourY}{-2.4}

\begin{document} 
\title[AMP with RBM priors]{Approximate Message Passing with Restricted Boltzmann Machine Priors}

\author{Eric~W.~Tramel$^{\dag}$, Ang{\'e}lique~Dr{\'e}meau$^{\cup}$, and Florent~Krzakala$^{\dag,\star}$}

\address{$\star$ Sorbonne Universit{\'e}s \&
				 UPMC Universit{\'e} Paris 06}
\address{$\dag$ Laboratoire de Physique Statistique (UMR 8550 CNRS),
			    {\'E}cole Normale Sup{\'e}rieure\\
			    ~~Rue Lhomond, 75005 Paris, France}
\address{$\cup$ ENSTA Bretagne \\
				~~~29806 Brest, France}

\begin{abstract} 
Approximate Message Passing (AMP) has been shown to be an excellent
statistical approach to signal inference and compressed sensing
problem. The AMP framework provides modularity in the choice of signal
prior; here we propose a hierarchical form of the Gauss-Bernouilli
prior which utilizes a Restricted Boltzmann Machine (RBM) trained on
the signal support to push reconstruction performance beyond that of
simple \iid priors for signals whose support can be well represented
by a trained binary RBM. We present and analyze two methods of RBM
factorization and demonstrate how these affect signal reconstruction
performance within our proposed algorithm. Finally, using the MNIST
handwritten digit dataset, we show experimentally that using an RBM
allows AMP to approach oracle-support performance.
\end{abstract} 

\newpage

\section{Introduction}
	\label{sec:intro}
Over the past decade, a groundswell in research has occurred in both
difficult inverse problems, such as those encountered in Compressed
Sensing (CS) \cite{CR2005a}, and in signal representation and
classification via deep networks. In recent years, Approximate Message
Passing (AMP) \cite{DMM2009} has been shown to be a near-optimal,
efficient, and extensible application of belief propagation to solving
inverse problems which admit a statistical description.

While AMP has enjoyed much success in solving problems for which an
\iid signal prior is known, only a few works have investigated the
application of AMP to more complex, structured priors. Utilizing such
complex priors is key to leveraging many of the advancements recently
seen in statistical signal representation. Techniques such as GrAMPA
\cite{BS2013} and Hybrid AMP \cite{rangan2011hybrid} have shown
promising results when incorporating correlation models directly
between the signal coefficients, and in fact the present contribution
is similar in spirit to Hydrid AMP.

Another possible approach is to not attempt to model the correlations
directly, but instead to utilize a bipartite construction via hidden
variables, as in the Restricted Boltzmann Machine (RBM)
\cite{Smo1986,Hin2002}. The RBM is an example of latent variable model, which
we distinguish from the fully visible models considered by \cite{BS2013,rangan2011hybrid}.
Such latent variable models can become quite powerful, as they 
admit interpretations such as feature extractors and multi-resolution
representations. 
As we will show, if a binary RBM can be trained to
model the support patterns of a given signal class, then the
statistical description of the RBM easily admits its use within
AMP. This is particularly interesting since RBMs are the building
blocks of \emph{deep belief networks} \cite{bengio2009learning} and have
enjoyed a surge of renewed interest over the past decade, partly due to 
the development of efficient training algorithms (e.g. Contrastive Divergence (CD)
\cite{Hin2002}). The present paper demonstrates the first
steps in incorporating deep learned priors into generalized linear
problems.


\section{Approximate message passing for compressed sensing}
	\label{sec:amp}
Loopy Belief Propagation (BP) is a powerful iterative message passing
algorithm for graphical models \cite{Pearl82,MM2009}. However, it
presents two main drawbacks when applied to highly-connected,
continuous-variable problems, as in CS:
first, the need to work with continuous probability distributions; and
second, the necessity to iterate over one such probability
distribution for each pair of variables. These problems can be
addressed by projecting the distributions onto their first two moments
and by approximating the messages, which exist on the edges of the factor graph
representation of the inference problem, by the marginals existing 
on the nodes of the factor graph. 
Applying both to BP for the CS problem, one obtains the AMP iteration.

AMP has been shown to be a very powerful algorithm for CS signal
recovery, especially in the Bayesian setting where an \emph{a priori} 
model of the unknown signal is given.  
In CS, one has the following forward model to obtain set
of $M$ observations $\mathbf{y}$,
\begin{equation}
y_{\eta} = \sum_{i=1}^N F_{\eta i} x_i + w_{\eta} \quad{\rm where}\quad 
    w_{\eta} \sim \mathcal{N}\p{0,\Delta},
\label{eq:forward_model}
\end{equation}
where $F$ is an $M\times N$ matrix,
for $M\ll N$, representing linear observations of an unknown signal
$\mathbf{x}$ which are then corrupted by an additive zero-mean white
Gaussian noise (AWGN) of variance $\Delta$. The measurement rate $\alpha = M/N$
is of particular interest in determining the difficulty of this inverse problem.
In the present work, we use subscript notation to denote the individual coefficients of vectors, i.e. $y_{\eta}$ refers to the $\eta^{th}$ coefficient of ${\bf y}$ 
and the double-subscript notation to refer to individual matrix
elements in row-column order.

We can use AMP to estimate a factorization, up to 
the first two moments, of the posterior distribution
  $P(\mathbf{x}|F,\mathbf{y}) \propto P_0(\mathbf{x})P(\mathbf{y}|F,\mathbf{x})$,
where $P(\mathbf{y}|F,\mathbf{x})$ is
the likelihood of an observation from the AWGN channel and
$P_0(\mathbf{x})$ is a prior on the signal. If the posterior is trivially factorized, or
if it can be approximated by a factorized distribution, one can estimate the unknown
signal by averaging over the posterior,
\begin{equation}
\hat{x}_i^{\rm MMSE} = \int{\rm d}x_i~~{x_i P(x_i | F, \mathbf{y})}, \quad \forall i,
\end{equation}
which is the minimum mean squared error (MMSE) estimate of $\mathbf{x}$. This is in
contrast to utilizing a maximum \textit{a posteriori} (MAP) approach
to the solution of this inverse problem. We refer the
reader to
\cite{DMM2009,DMM2010,Ran2011,VS2013,KrzakalaPRX2012},
and in particular to \cite{KMS2012}, for the present notation and the
derivation of AMP from BP and now give directly the iterative form of the
algorithm. Given an estimate of the factorized posterior mean
$a_i$ and variance $c_i$ for each element of $\mathbf{x}$, a single step of AMP
iteration reads
\begin{eqnarray}
V^{t+1}_\eta &=& \sum_i F_{\eta i}^{2} c_i^t \, ,\label{eq:V} \\
\omega^{t+1}_\eta  &=& \sum_i F_{\eta i} a^t_i -
(y_\eta-\omega^t_\eta)\frac{V^{t+1}_\eta}{\Delta+V^{t}_\eta} \, , \label{eq:omega}\\
(\Sigma^{t+1}_i)^2 &=& \left[ \sum_\eta \frac{F^2_{\eta
      i}}{\Delta+V^{t+1}_\eta} \right]^{-1} \, ,  \label{eq:S}\\
R^{t+1}_i &=& a^t_i + (\Sigma^{t+1}_i)^2 \sum_\eta F_{\eta i}
\frac{(y_\eta - \omega^{t+1}_\eta)}{\Delta+V^{t+1}_\eta} \, .\label{eq:R}
\end{eqnarray}
These terms can be interpreted as follows. The variables 
$\left\{\omega_\eta\right\}$ and $\left\{V_\eta\right\}$ represent
the first and second moments, respectively,
of the marginalized messages on the $M$ factors of the factor graph. 
The variables $\left\{R_i\right\}$ and $\left\{\Sigma_i^2\right\}$ 
represent the first and second moments, respectively, of the marginalized
messages on the $N$ signal variables of the factor graph. As such,
these represent the \emph{AMP field} on the signal, which are the
observational beliefs about the continuous factorized PDF at
each signal variable. To find the final factorized posterior of the signal,
up to the two moment approximation, we must augment this 
AMP field by the signal prior. Once the new values of $R_i$ and $\Sigma_i^2$ 
are computed, the new
estimates of the posterior mean and variance, $a_i$ and $c_i$, 
given the prior $P_0(\mathbf{x})$, are calculated as
\begin{eqnarray}
  a_i^{t+1}  &\defined \int {\rm d}x_i~~\frac{x_i}{\Zi} P_0(x_i)\mathcal{N}(x_i;R_i,\Sigma_i^2), \label{eq:fa} \\
  c_i^{t+1} &\defined \int {\rm d}x_i~~\frac{x_i^2}{\Zi} P_0(x_i)\mathcal{N}(x_i;R_i,\Sigma_i^2) - \p{a_i^{t+1}}^2, 
  \label{eq:fc}
\end{eqnarray}
where $\Zi = \int{\rm d}x_i~~P_0(x_i)\mathcal{N}(x_i;R_i,\Sigma_i^2)$ 
is a normalization constant, commonly referred to as 
a \emph{partition function} in statistical physics.  
The final form of these equations for differing values of $P_0(x_i)$ 
are given in \cite{KMS2012}.
As seen in
Eqs. (\ref{eq:fa}) and (\ref{eq:fc}), in order to use the AMP
framework, one must know some information about the class of signals
from which $\mathbf{x}$ is drawn. Commonly in CS, we are interested in
the case where the signal $\mathbf{x}$ is
\emph{sparse}, that is, very few of its coefficients are non-zero. The concept
of sparsity implies an assumption that the amount of information
required to represent $\mathbf{x}$ is actually far less than its dimensionality
would admit. Essentially, a sparse prior on $\mathbf{x}$ is extremely informative
due to its low-entropic nature.

Much of the CS literature focuses on convex approaches to this inverse
problem.  And so, a convex $\ell_1$ norm is used as a regularizer to
bias solutions to the inverse problem towards sparsity. Within the
probabilistic framework, this corresponds to a selection of a Laplace
distribution for $P_0(x_i)$.

However, AMP is not restricted to convex priors,
but can utilize non-convex priors of arbitrary complexity. For
example, one can use a two-mode Gauss-Bernoulli (GB) prior to model sparse
signals (as was considered in detail
in \cite{Ran2011,VS2013,KrzakalaPRX2012}) such as
\begin{equation}
  P_0(\mathbf{x})
  = \prod_i   P_0(x_i)
  = \prod_i   \sum_{v_i\in\bra{0,1}} P_0(v_i)P_0(x_i|v_i),
\end{equation}
where we have introduced Bernoulli random variables $v_i$ such that $\expval{v_i} = \rho$,
on which the values $\bra{x_i}$ are conditioned as 
$P_0(x_i|v_i=1) = \mathcal{N}(\mu,\sigma^2)$, 
and $P_0(x_i|v_i=0) = \delta(x_i)$, where
$\delta(\cdot)$ is the zero-centered Dirac delta distribution.
This leads to the final expression for the prior,
\begin{equation}
  P_0(\mathbf{x};\rho) = \prod_i\left( (1-\rho)\delta\p{x_i} + \frac{\rho}{\sqrt{2\pi\sigma^2}} e^{-\frac{(x_i-\mu)^2}{2\sigma^2}}\right).
\label{GB}
\end{equation}
The GB prior has two possible modes: a zero mode
and a non-zero one. If we denote the set of non-zero coefficients of
$\mathbf{x}$, those for which $v_i = 1$, 
to be the \textit{support}, then the GB prior models both
the on- and off-support probability.  For $\rho \in [0,1]$, the
distribution splits the probabilistic weight between a hard
(deterministic) constraint of $x_i = 0$ and a normal distribution for
arbitrary parameters that model the on-support coefficients. For a
fixed on-support distribution, here for fixed $\mu$ and $\sigma^2$,
the value of $\rho$ controls how informative $P_0(\mathbf{x};\rho)$
is. The more informative this prior, the fewer measurements, smaller
$\alpha$, are required in order to successfully infer $\mathbf{x}$
from $\mathbf{y}$. Of course, if $\mathbf{x}$ is not truly drawn from
$P_0(\mathbf{x};\rho)$, more measurements will be required to account
for the mismatch between the true signal and the assumed signal model,
as is the case with signals which are not truly sparse but merely
compressible.

If, rather than a fixed probability of being on-support for all sites,
we instead have a local probability for each specific site to be
on-support, we can write an independent, but non-identically distributed, prior. 
For conciseness, we refer to this property as ``non-\iid'' in the 
remainder of this paper.
We easily generalize
Eq. (\ref{GB}) to
\begin{equation}
  P_0(\mathbf{x};\bra{\rho_i})
  = \prod_i \left( (1-\rho_i)\delta\p{x_i} + \frac{\rho_i}{\sqrt{2\pi\sigma^2}}
      e^{-\frac{(x_i-\mu)^2}{2\sigma^2}} \right).
\end{equation}

This change in the prior must also be reflected in the computation of the means and variances used in in Eqs. (\ref{eq:fa}) and (\ref{eq:fc}). In fact, the partition function becomes
\begin{eqnarray}
  \Zi 
    &= (1-\rho_i) \frac{1}{\sqrt{2\pi\si}}\bexp{-\frac{\ri^2}{2\si}} \nn\\
    &\quad + \rho_i \frac{1}{\sqrt{2\pi\p{\si +\sigma^2}}}\bexp{-\frac{(\ri-\mu)^2}{2(\si+\sigma^2)}}, \nn\\
    &= (1-\rho_i) \Ziz + \rho_i \Zinz,
\end{eqnarray}
where we can write the partition using two sub-partition terms
related to the off-support ($\Ziz$) and on-support ($\Zinz$) probabilities. From this
partition function, for a given setting of $\ri$ and $\si$ (which result from the AMP
evolution), we can write the posterior means and variances, $\mathbf{a}$ and $\mathbf{c}$
according to the fixed GB prior parameters $\bra{\rho_i}$, $\mu$, and $\sigma^2$.

A nice feature of this two-mode prior is that it also admits a natural estimation of
the probability
of a particular coefficient to be on- or off-support. Specifically,
at a given point in the AMP evolution, we have
$\ampPnz = \frac{\rho_i\Zinz}{\Zi}$ and $\ampPz = \frac{(1-\rho_i)\Ziz}{\Zi}$, leading to
\begin{eqnarray}
  P^{\amp}_i(v_i) 
    &= \p{\frac{(1-\rho_i)\Zinz}{\Zi}}^{v_i}\p{\frac{\rho_i\Ziz}{\Zi}}^{1-v_i},\nn\\
    &\propto \bexp{v_i\ln\frac{\Zinz}{\Ziz} + v_i \ln\p{\frac{\rho_i}{1-\rho_i}}},\nn\\
    &\propto e^{v_i \left( \ln\gi+ \ln {\frac{\rho_i}{1-\rho_i}}\right)},
    \label{eq:amp_field}
\end{eqnarray}
where
$\gi = \frac{\Zinz}{\Ziz}$ has the natural logarithm
\begin{eqnarray}
  \ln\gi 
  &= \frac{\ri^2}{2\si} - \frac{(\ri-\mu)^2}{2(\si+\sig^2)}
  + \ln\sqrt{\frac{\si}{\si+\sig^2}}. 
\label{eq:log_gi}
\end{eqnarray}
The additional
information provided by AMP results in a
modified support probability 
$\tilde{P}(\mathbf{v})\propto P(\mathbf{v}) \prod_i e^{v_i \ln \gi}$. 
Explicating 
$P(\mathbf{v})$
allows us to envision more complex support models 
for the coefficients of $\mathbf{x}$. The previous model
assumes the independence between the coefficients of $\mathbf{x}$, however, the 
existence of dependencies, now well-acknowledged for many natural signals
as \emph{structured sparsity}, can be leveraged through joint models. 
In the variational Bayesian context, we cite
\cite{rangan2011hybrid,schniter2010turbo,dremeau2012boltzmann}, which
consider neighborhood probabilities, Markov chains, and so-called 
\emph{Boltzmann machines}, respectively, as generic support models.
In similar vein, we propose the use of a binary 
RBM as a joint support model. In contrast with the models of 
\cite{rangan2011hybrid,schniter2010turbo}, the RBM model can provide
a more accurate modeling of support correlations. Also, the RBM model can
be trained at low cost \cite{Hin2002}, which is the main bottleneck of the \emph{general}
Boltzmann machine model used in \cite{dremeau2012boltzmann}.

\section{Binary restricted Boltzmann machines}
	\label{sec:rbm}
An RBM is an energy based model
defined over both visible and a set of latent, or \emph{hidden}, variables. 
From the perspective of statistical physics, the RBM can be 
viewed as a boolean Ising model existing on a bipartite graph.
The joint probability distribution over the visible and 
hidden layers for the RBM is given by
\begin{equation}
  P(\vis,\h) \propto e^{-E(\vis,\h)}\,,
\end{equation}
where the energy $E(\vis,\h)$ reads
\begin{equation}
E(\vis,\h) =  -\sum_i \vbiasi\visi - \sum_j \hbiasj\hj -\sum_{i,j}
\Wij\visi\hj\,.
\end{equation}
The RBM model is described by the two sets of biasing coefficients
$\vbias$ and $\hbias$ on the visible and hidden layers, respectively,
and the learned connections between the layers represented by the
matrix $\rbmW$.

In the sequel, we leverage the connection between the RBM and the well
known results of statistical physics to
discuss a simplification of the RBM under the so-called
mean-field approximation in both the $1^{\rm st}$-order
and $2^{\rm nd}$-order approximation, known as
Thouless-Anderson-Palmer (TAP) approximation \cite{ThoulessAnderson77,mezard:87,Yedidia01},
in order to obtain factorizations over the visible and hidden layers from
this joint distribution.

\subsection{Mean-field approximation of the RBM}
Given the value of an Energy-function $E(\{{\bf x\}})$, also called
Hamiltonian in physics, a standard technique is to use the Gibbs
variational approach where the Gibbs free energy $\mathcal{F}$ is
minimized over a trial distribution, $P_{\rm var}$, with
\begin{eqnarray}
  \mathcal{F}(\{{P_{\rm var}\}}) = \left< E(\{{\bf x\}})\right>_{\{{P_{\rm var}\}}} - S_{\rm Gibbs}\left(P_{\rm var}\right)
\end{eqnarray}
where $\left<~\cdot~\right>_{\{{P_{\rm var}\}}}$ denotes the average
over distribution $P_{\rm var}$ and $S_{\rm Gibbs}$ is the Gibbs
entropy.

It is instructive to first review the simplest variational solution,
namely, the $1^{\rm st}$-order na{\"i}ve mean-field (NMF) approximation, 
where $P_{\rm var}=\prod_i Q_i(x_i)$. Within this ansatz, a classical computation shows that the
free energy, in the case of the binary RBM, reads as
\begin{eqnarray}
\Fmfa^{\rm RBM}\p{\mvis,\mh}
  &=  - \sum_i \vbiasi\mvisi   -\sum_j \hbiasj\mhj  \nn-\sum_{\left<i,j\right>}\Wij\mvisi\mhj \nn
\\
  &\quad +   \sum_i \s{ \ent{\mvisi} + \ent{1 - \mvisi} }  \nn\\
  &\quad +\sum_j \s{ \ent{\mhj}   + \ent{1 - \mhj}},
\end{eqnarray}
where $H(x)=x \ln(x)$ and 
$\mvisi = \expval{\visi} = \prob{\visi = 1}$.
Since the
hidden variables are also binary, this identity is
equally true for $\mhj$.

The fixed points of the means of both the visible and hidden 
units are of particular interest.
With these fixed points 
we can calculate a factorization for both $P(\vis)$ and $P(\h)$.
First, we look at the derivatives
of the NMF free energy for the RBM w.r.t. the visible and hidden
sites, which, when evaluated at the critical point, gives us the
fixed-point conditions for the expected values of the variables
\begin{eqnarray}
  \mvisi &= \sigm(\vbiasi + \sum_j\Wij\mhj),\label{eq:rbmmf_v}\\
  \mhj   &= \sigm(\hbiasj + \sum_i\Wij\mvisi),\label{eq:rbmmf_h}
\end{eqnarray}
where $\sigm\p{x} \defas [1 + e^{-x}]^{-1}.$ These equations are in line with the assumed NMF
fixed-point conditions used for finding the ``site activations'' given
in the RBM literature.  In fact, they are often used as 
a fixed-point iteration (FPI) to find the minimum free energy.

We will now modify the NMF solution via a $2^{\rm nd}$-order correction,
as was originally shown for RBMs in \cite{GTZ2015} using TAP.
The TAP approach is a classical tool in
statistical physics and spin glass theory which
improves on the NMF approximation by taking into account further correlations. 
In many situations, the improvement is drastic, making the TAP
approach very popular for statistical inference \cite{Yedidia01}.
There are many ways in which the TAP equations can be presented. We
shall refer, for the sake of this presentation, to the known
results of the statistical physics community. One approach to derive
the TAP approximation is to recognize that the NMF free energy is
merely the first term in a perturbative expansion in power of the
coupling constants $\rbmW$, as was shown by Plefka
\cite{plefka1982convergence}, and to keep the $2^{\rm nd}$-order
term. Alternatively, one may start with the Bethe approximation and use
the fact that the system is densely connected
\cite{Yedidia01,mezard:87}. Proceeding according to Plefka, 
the TAP free energy reads
\begin{eqnarray}
    \Ftap^{\rm RBM}\p{\mvis,\mh}=   \Fmfa^{\rm RBM} \p{\mvis,\mh} 
- \frac{1}{2} \sum_{\edges}\Wij^2\vvisi\vhj,
\label{eq:tapfe}
\end{eqnarray}
where we have denoted the variances of hidden and visible variables,
$\vh$ and $\vvis$, respectively, as
$\vhj   = \expval{\hj^2} - \expval{\hj}^2 = \mhj - \mhj^2$ 
and
$\vvisi = \expval{\visi^2} - \expval{\visi}^2 =\mvisi - \mvisi^2$.
Repeating the extremisation, one now finds
\begin{eqnarray}
    \mvisi &= \sigm\left(\vbiasi +\sum_j\Wij\mhj 
                       + \p{\frac{1}{2} -
                 \mvisi}\sum_j\Wij^2\vhj\right),\label{eq:rbmtap_v}\\
    \mhj  &= \sigm\left(\hbiasj + \sum_i\Wij\mvisi + \p{\frac{1}{2} -
                \mhj}\sum_i\Wij^2\vvisi\right).\label{eq:rbmtap_h}
\end{eqnarray}

Eqs. (\ref{eq:rbmtap_v}) and (\ref{eq:rbmtap_h}), 
often called the TAP equations, can be seen are an
extension of the mean field iteration of Eqs. (\ref{eq:rbmmf_v}) and (\ref{eq:rbmmf_h}). 
The additional term is called the Onsager retro-action term in statistical
physics \cite{mezard:87}. In fact, these are the tools one uses in
order to derive AMP itself. Given an RBM model, we now have two approximated
solutions to obtain the equilibrium marginal through the iteration of
either Eqs. (\ref{eq:rbmmf_v}) and (\ref{eq:rbmmf_h}) or Eqs. (\ref{eq:rbmtap_v}) and 
(\ref{eq:rbmtap_h}). 

Because of the sigmoid functions in the fixed-point conditions, 
iterating on these fixed points will not wildly diverge. 
However, it is possible that such an FPI will 
arrive at one of the two trivial solutions for the factorization, either the
ground-state of the field-less RBM or $\frac{1}{2}({\rm sign}(\vbias) + 1)$. 
Whether or not the FPI arrives at these trivial points 
relies on the balance between the evolution of the FPI and the contributing fields.
It may also enter an oscillatory state, especially if the learned RBM couplings are too
large in magnitude, reducing the accuracy of the Plefka expansion which assumes small
magnitude couplings. We shall see, however, that this FPI works extremely well in practice.

\section{RBMs for AMP}
	\label{sec:rbmamp_all}

\begin{figure}\begin{center} 
        \begin{tikzpicture}[auto,thick, node distance = 3em, >=triangle 45,scale=1]

\draw 
    node at (3,-0.5)[factor] (y1) {}
    node at (3,-1.5)[factor] (y2) {}
    node at (3,-2.5)[factor] (y3) {};

\draw 
    node at (1, 0)[var] (x1) {}
    node at (1,-1)[var] (x2) {}
    node at (1,-2)[var] (x3) {}
    node at (1,-3)[var] (x4) {};

\draw 
    node at (0, 0)[factor] (p1) {}
    node at (0,-1)[factor] (p2) {}
    node at (0,-2)[factor] (p3) {}
    node at (0,-3)[factor] (p4) {};

\draw 
    node at (-1, 0)[var] (v1) {}
    node at (-1,-1)[var] (v2) {}
    node at (-1,-2)[var] (v3) {}
    node at (-1,-3)[var] (v4) {};

\draw 
    node at (-3,-0.5)[var] (h1) {}
    node at (-3,-1.5)[var] (h2) {}
    node at (-3,-2.5)[var] (h3) {};

\draw
    (y1) -- (x1)
    (y1) -- (x2)
    (y1) -- (x3)
    (y1) -- (x4)
    (y2) -- (x1)
    (y2) -- (x2)
    (y2) -- (x3)
    (y2) -- (x4)
    (y3) -- (x1)
    (y3) -- (x2)
    (y3) -- (x3)
    (y3) -- (x4);

\draw
    (h1) -- (v1)
    (h1) -- (v2)
    (h1) -- (v3)
    (h1) -- (v4)
    (h2) -- (v1)
    (h2) -- (v2)
    (h2) -- (v3)
    (h2) -- (v4)
    (h3) -- (v1)
    (h3) -- (v2)
    (h3) -- (v3)
    (h3) -- (v4);

\draw
    (x1) -- (v1)
    (x2) -- (v2)
    (x3) -- (v3)
    (x4) -- (v4);

\draw
    node at ( 3,-4) (obslabel) {$\mathbf{y}$}
    node at ( 2,-4) (projlable) {$F$}
    node at ( 1,-4) (varlabel) {$\mathbf{x}$}
    node at (-1,-4) (vislabel) {$\mathbf{v}$}
    node at (-2,-4) (rbmlabel) {$\mathbf{W}$}
    node at (-3,-4) (hidlabel) {$\mathbf{h}$}
    node at (-2, 1) (priorlabel) {Prior Side}
    node at ( 2, 1) (priorlabel) {Observation Side};

\draw [decorate,decoration={brace,amplitude=0.5em,raise=0.5em},yshift=0em]
    (-3,0.3) -- (-1,0.3);
\draw [decorate,decoration={brace,amplitude=0.5em,raise=0.5em},yshift=0em]
    ( 1,0.3) -- ( 3,0.3);

\end{tikzpicture} 
    \end{center} \caption{\label{fig:factorgraph}
        Graphical representation of the statistical dependencies of
        the proposed RBM-AMP: The right side represents observational side, 
        with linear constraints, while the
        left side represent the RBM prior on
        the support of the signal (see text).} 
    \end{figure}
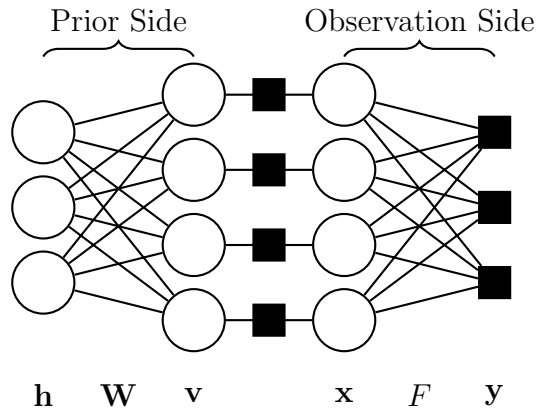

The scope of this work is to use the RBM model within the AMP
framework and to perform inference according to the graphical model
depicted in Fig. \ref{fig:factorgraph}. Here, we would like to utilize the
binary RBM to give us information on each site's likelihood of being
\emph{on-support}, that is, its probability to be a non-zero
coefficient. This shoehorns nicely into our sparse GB prior as in
Eq. (\ref{GB}).

Since, in the case of an RBM, $P(v_i)$ has the classical exponential
form of an energy-based model, we see from Eq. (\ref{eq:amp_field})
that the information provided by AMP simply amounts to an
additional local bias on the visible variables equal to $\ln\gi$. That
is, the AMP-modified RBM free energy is simply the introduction of an
additional field term along with a constant bias,
\begin{eqnarray}
\FErbmamp &= \mathcal{F}^{\rm RBM}\p{\mvis,\mh} - \sum_i \mvisi\ln\gi+\mathcal{C}\,.
\end{eqnarray}

As we can see, this field effect only exists on the visible layer of the RBM.
Because of this, the AMP framework does not put any extra influence on the 
hidden layer, but only on the visible layer. Thus, the 
fixed point of the hidden layer means is not influenced by AMP, but the
visible ones are. With respect to (\ref{eq:rbmmf_v}) and
(\ref{eq:rbmtap_v}), the AMP-modified fixed points of the the visible
variable means contain one additional additive term within the sigmoid,
giving
\begin{eqnarray}
	\mvisi &= \sigm\p{\vbiasi + \ln\gi + \sum_j\Wij\mhj},\label{eq:rbmampmf_v}
\end{eqnarray}
for the NMF-based fixed point. For the TAP we have
\begin{eqnarray}
	\mvisi 
        = \sigm\p{\vbiasi + \ln\gi + \sum_j\Wij\mhj 
        + \p{\frac{1}{2} - \mvisi}\sum_j\Wij^2\vhj}.
   \label{eq:rbmamptap_v}
\end{eqnarray}



The most direct approach for factorizing 
the AMP influenced RBM is to construct a fixed-point iteration (FPI) 
using the NMF or TAP fixed-point conditions.
We give the final construction of the AMP algorithm with the
RBM support prior in Alg. \ref{alg:rbmamp} using this approach. This approach can
be understood in terms of its graphical representation given in Fig. \ref{fig:factorgraph},
which shows the network of statistical dependencies from the observations to the hidden
RBM hidden units.
Given some initial condition for $\mvis$ and $\mh$, we can successively estimate the
visible and hidden layers via their respective fixed-point equations. Empirically, we have
found that the hidden-layer variables should be initialized to zero, while the visible
side variables can be initialized to zero, uniformly randomly, or by drawing a random
initial condition according to the distribution implied by the RBM visible bias $\vbias$.

For the FPI schedule, one might intuitively attempt an approach 
similar to persistent contrastive divergence \cite{Tie2008} and allow the values of
$\mvis$ and $\mh$ persist throughout the AMP FPI, taking only a single FPI step on the
RBM magnetizations at each AMP iteration. Indeed, we have found this to be 
the most computationally efficient integration of the RBM into the AMP framework. However,
we point out that this persistent strategy should only begin after a few AMP iterations have
been completed. In the first iterations, the RBM factorization should instead be allowed
to converge on a value of $\mvis$. Persistence of the hidden and visible magnetizations 
from the first iteration leads to poor reconstruction performance, especially when 
$\alpha$ is small. Because the early values of $\{\ln\gi\}$ can be quite weak, using 
only a single update step on the RBM magnetizations early in the reconstruction
does not adequately enforce the support prior. Later, as the AMP fields grow in magnitude, 
a single-step persistent update of the RBM magnetizations can be used to decrease run time.

Finally, after obtaining the value of $\mvis$, either by running the RBM FPI until 
convergence or by taking a single step on $\mvis$, we must infer the correct values
$\{\rho_i\}$. One might at first attempt to use the setting $\{\rho_i = \mvisi \}$,
however, this is an improper approach for the hierarchical support prior we have 
proposed. Instead, one should use
\begin{eqnarray}
    \rho_i 
        &= \frac{\mvisi \ampZiz}{\mvisi \ampZiz + (1-\mvisi) \ampZinz}
        = {\rm sigm}\p{\ln\mvisi - \ln(1-\mvisi) - \ln\gi},
    \label{eq:rho_update}
\end{eqnarray}
to obtain the correct per-pixel sparsity terms to use in conjunction with the 
standard Gauss-Bernoulli form of (\ref{eq:fa}) and (\ref{eq:fc}).




\begin{algorithm}[H]
   	\caption{AMP with RBM Support Prior\label{alg:rbmamp}}  
	\begin{algorithmic}
		\STATE {\bfseries Input:} $F$, $\mathbf{y}$, $\mathbf{W}$, $\vbias$, $\hbias$, PersistentStart
		\STATE \emph{Initialize}: $\mathbf{a}$,$\mathbf{c}$,$\mvis$,
														  $\{\mhj = 0,~\forall j\}$, $\{\vhj = 0,~\forall j\}$,
														  ${\rm Iter} =1$
		\REPEAT		
			\STATE AMP Update on $\{V_{\eta},\omega_{\eta}\}$ 
			       via (\ref{eq:V}), (\ref{eq:omega})
			\STATE AMP Update on $\{\ri,\si\}$ 
			       via (\ref{eq:R}), (\ref{eq:S}) 
			\STATE Calculate $\{\ln\gi\}$ via (\ref{eq:log_gi})~~$\forall i$
			\IF{Iter $<$ PersistentStart}
				\STATE \emph{(Re)Initialize}: $\{\mhj,\vhj = 0,~\forall j\}$
				\STATE \emph{(Re)Initialize}: $\{\mvisi = 0,~\forall i\}$
				\REPEAT
					\STATE Update $\{\mhj,\vhj\}$ via (\ref{eq:rbmmf_h}) or (\ref{eq:rbmtap_h})
					\STATE Update $\{\mvisi,\vvisi\}$ via (\ref{eq:rbmampmf_v}) or (\ref{eq:rbmamptap_v})
				\UNTIL{Convergence on $\mvis$}
			\ELSE
					\STATE Update $\{\mhj,\vhj\}$ via (\ref{eq:rbmmf_h}) or (\ref{eq:rbmtap_h})
					\STATE Update $\{\mvisi,\vvisi\}$ via (\ref{eq:rbmampmf_v}) or (\ref{eq:rbmamptap_v})
			\ENDIF
			\STATE Calculate $\{\rho_i\}$ via (\ref{eq:rho_update})
			\STATE AMP Update on $\{a_i\}$ using $\{\rho_i\}$ 
				   via (\ref{eq:fa}) 
			\STATE AMP Update on $\{c_i\}$ using $\{\rho_i\}$
				   via (\ref{eq:fc})
			\STATE ${\rm Iter} \gets {\rm Iter} + 1$
		\UNTIL{Convergence on $\mathbf{a}$}
	\end{algorithmic}
\end{algorithm}



\section{Numerical results with MNIST}
	\label{sec:results}
To show the efficacy of the proposed RBM-based support prior within
the AMP framework, we present a series of experiments using the 
MNIST database of handwritten digits. Each sample of the database
is a $28\times28$ pixel image of a digit with values in the range
$[0,1]$. In order to build a binary RBM model of the support of these handwritten digits,
the training data was thresholded so that all non-zero pixels were given a 
value of $1$. We then train an RBM with 500 hidden units on 60,000 training 
samples from the binarized MNIST training set using the sampling-based 
contrastive divergence (CD-1) technique \cite{Hin2002}
for 100 epochs, averaging the RBM model parameter gradients across 100-sample minibatches,
at a learning rate of 0.005 under the prescriptions of \cite{Hin2010}.
We additionally impose an $\ell_2$ weight-decay penalty on the magnitude of the elements of
$\rbmW$
at a strength of 0.001. Such penlaties, while also desirable for learning performance 
\cite{Hin2010}, are necessary for the TAP-based AMP-RBM due to the fact that the Plefka 
expansion of Eq. (\ref{eq:tapfe}) is reliant on the magnitudes of $\rbmW$ being small.
For further reading on training RBMs, as well as other undirected graphical models, we
refer the reader to \cite{Fis2014}.
Once the generative RBM support model is obtained, we construct the CS experiments as follows.
For a given measurement rate $\alpha$, we draw the \iid entries of $F$ from
a normal distribution of variance $1/\sqrt{N}$. The linear projections
$F\vx$ are 
subsequently corrupted with an AWGN of variance $\Delta = 10^{-8}$ to form $\vy$. In all 
experiments we utilize the first 300 digit images from the MNIST test set. 
\begin{figure}
\centering
  \includegraphics[width=0.45\textwidth]{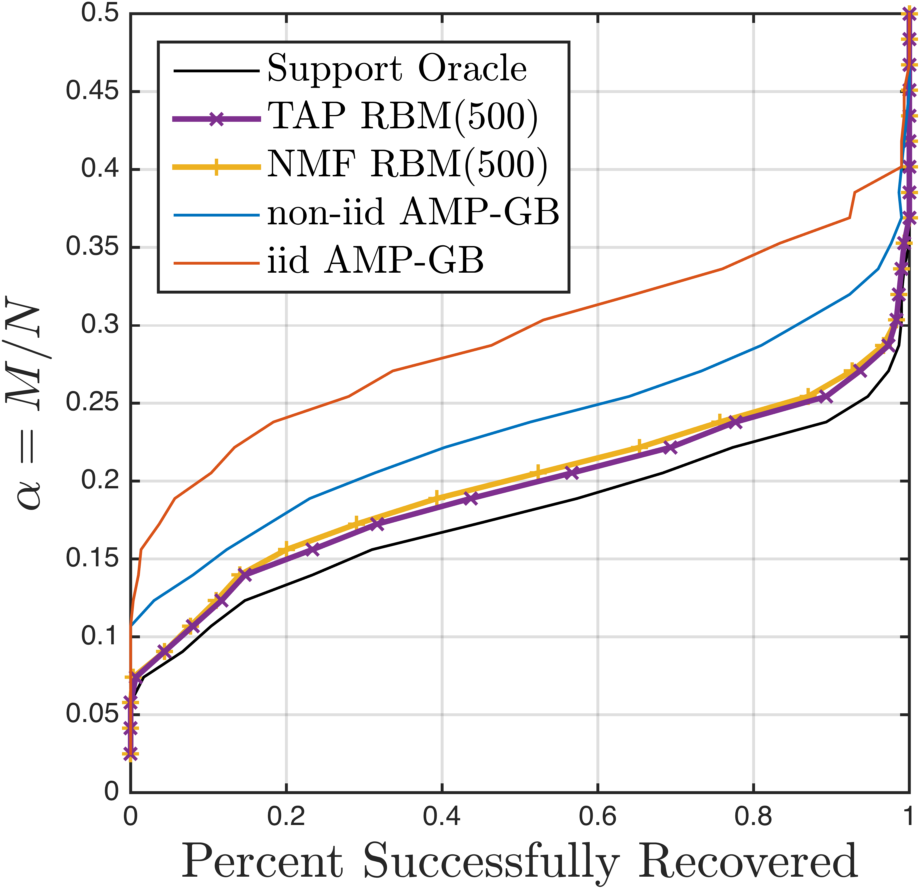}
  ~~~
  \includegraphics[width=0.45\textwidth]{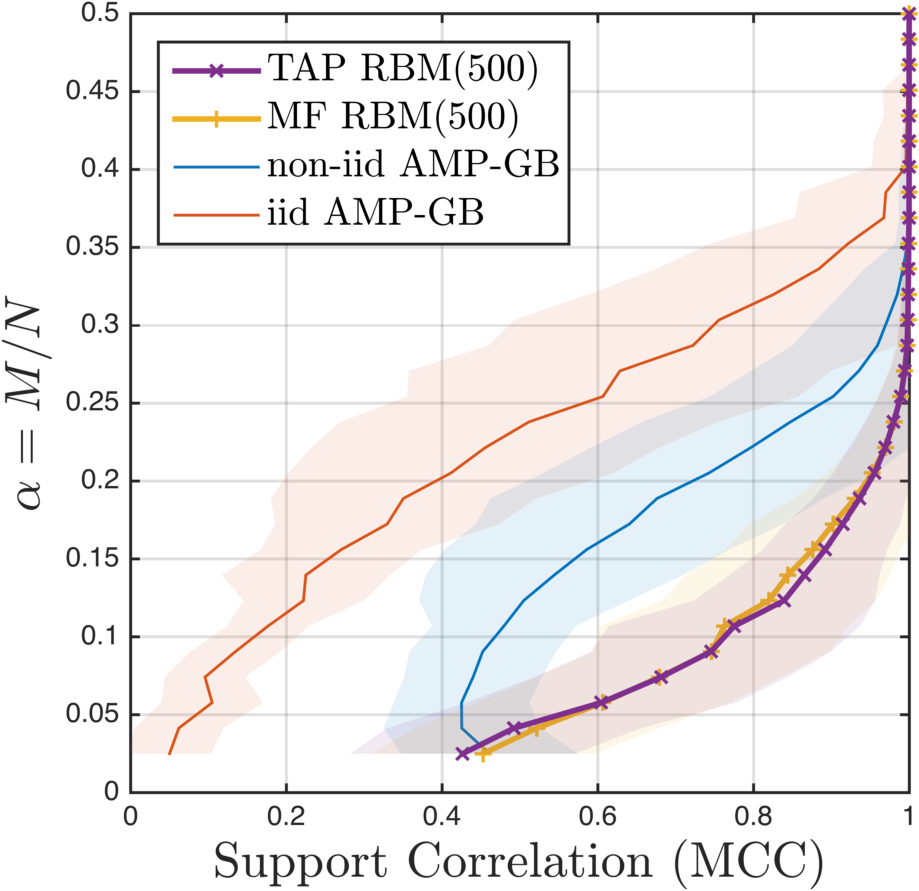}
\caption{\label{fig:results}    
Comparison of MNIST reconstruction performance over 300 test set digits
over the measurement rate $\alpha$. For both charts, lower curves 
represent greater reconstruction performance.
\textbf{(Left)}    
    Percentage of test set digit images successfully (MSE $\leq 10^{-4}$) 
    recovered. Note that the AMP-RBM model comes very close to the oracle
    reconstruction performance bound for this test set.
\textbf{(Right)}
    Correlation, in terms of MCC, 
    between the recovered support and the true digit image support. Here,
    the solid lines represent the average correlation at each tested $\alpha$,
    while the solid regions represent the range of one standard deviation of
    the correlations over the test set.}
\end{figure}

\begin{figure}
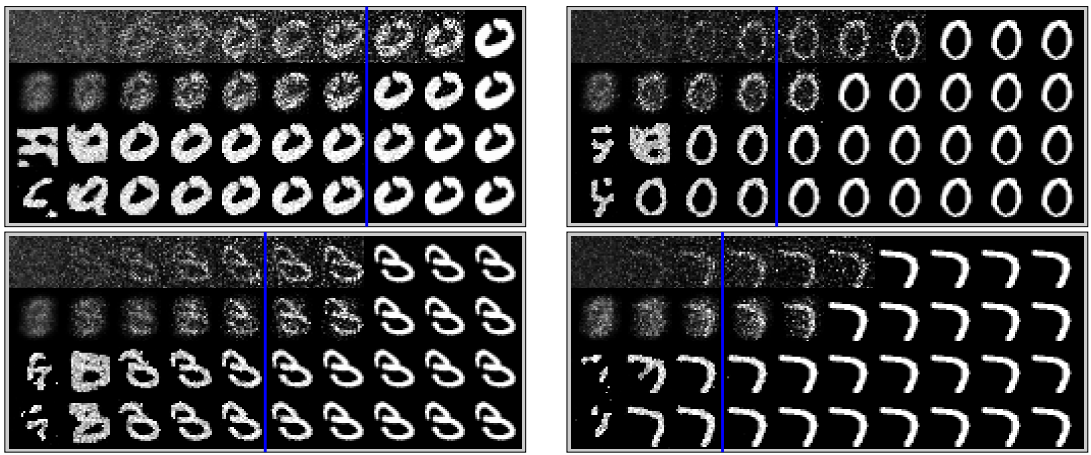

\centering
  \begin{minipage}{0.45\textwidth}
    \centering
    \raisebox{0.0in}{\resizebox{\textwidth}{0.420\textwidth}{\input{fig_recoveryCompare}}}\\
    \raisebox{2.80in}{\resizebox{\textwidth}{0.420\textwidth}{\input{fig_recoveryCompareC}}}
  \end{minipage} 
  ~
  \begin{minipage}{0.45\textwidth}
    \centering
    \raisebox{0.0in}{\resizebox{\textwidth}{0.420\textwidth}{\input{fig_recoveryCompareB}}}\\
    \raisebox{2.80in}{\resizebox{\textwidth}{0.420\textwidth}{\input{fig_recoveryCompareD}}}
  \end{minipage} 
  \vspace{-2.60in}
\caption{\label{fig:vis_results}    
  Visual comparison of reconstructions for four test digits across
  $\alpha$ for the same experimental settings. The rows of each box, 
  from top to bottom, correspond to the reconstructions providied by \iid AMP-GB,
  non-\iid AMP-GB, the proposed approach with NMF RBM factorization,
  and the proposed approach with TAP RBM factorization,
  respectively. The columns of each box, from left to right, 
  represent the values
  $\alpha = 0.025, 0.074, 0.123, 0.172, 0.222, 0.271, 0.320, 0.369, 0.418,
  0.467$. The advantages provided by the proposed approach are
  clearly seen by comparing the last row to the first one. The
  digits shown have $\rho = 0.342$ (top left), $\rho = 0.268$ (bottom left),
  $\rho = 0.214$ (top right), and $\rho = 0.162$ (bottom right). 
  The vertical blue line represents the $\alpha = \rho$ 
  oracle exact-reconstruction boundary for each reconstruction task.}
\end{figure}

We compare the following approaches in Fig. \ref{fig:results}. First, we show the
reconstruction performance of AMP using an \iid GB prior (AMP-GB), assuming that the 
true image-wide empirical $\rho$
is given as a parameter for each specific test image. 
Next, we demonstrate a simple modification to
this procedure: the GB prior is assumed to be non-\iid and the values of $\bra{\rho_i}$
are empirically estimated from the training samples as the probability of each
pixel to be non-zero. We expect that our proposed approach should at least perform
as well as non-\iid AMP-GB, as this same information should be encoded in 
$\vbias$ for a properly trained RBM model. 
Hence, this approach should correspond to
an RBM with zero couplings. We also show the performance of the proposed approach: 
AMP used in conjunction with the RBM support model. We present results for both the
NMF and TAP factorizations of the RBM. For the AMP-RBM approaches, we start persistence
after AMP 50 iterations. 
For all tested approaches, we assume that $\Delta$ is known
to the reconstruction algorithms, as well as the prior parameters $\mu$ and $\sigma^2$.
This is not a strict requirement, however, as channel and prior parameters can be 
estimated within the AMP iteration if so desired \cite{KMS2012}.

  \begin{figure}
  \centering
  \includegraphics[width=0.45\textwidth]{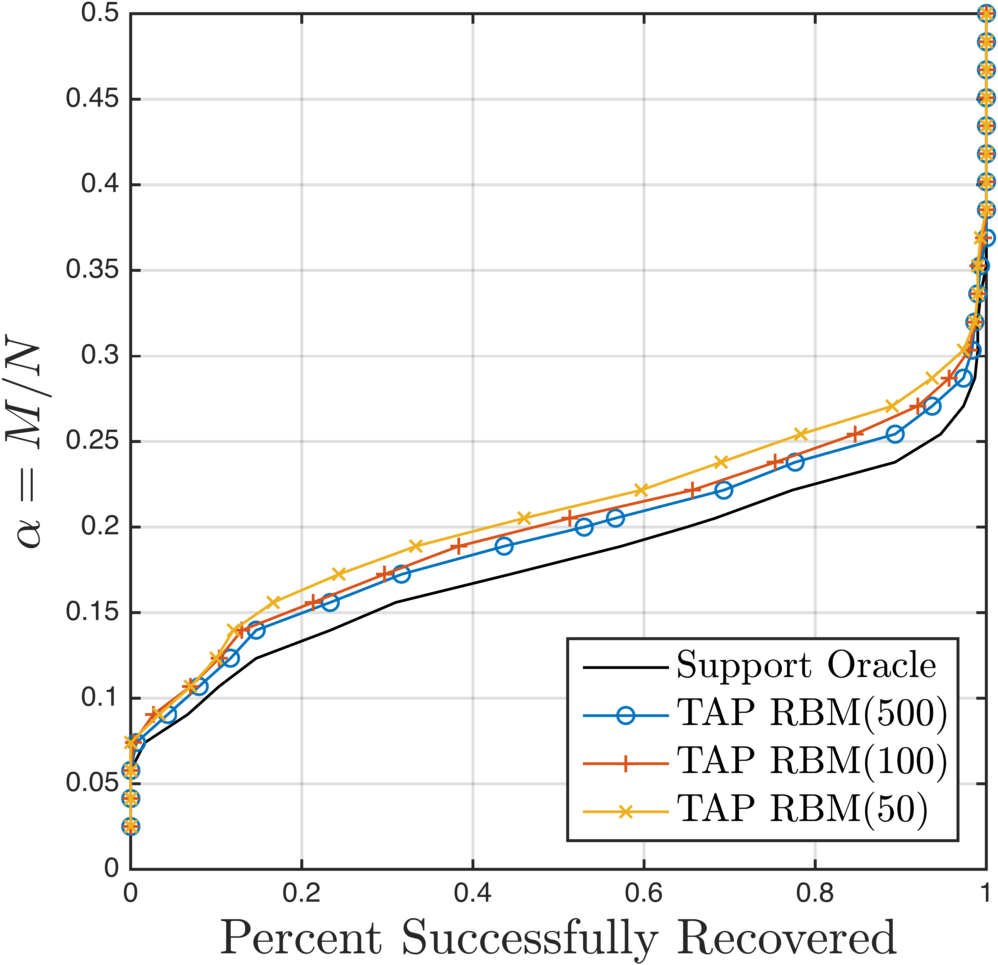}\quad
  \includegraphics[width=0.45\textwidth]{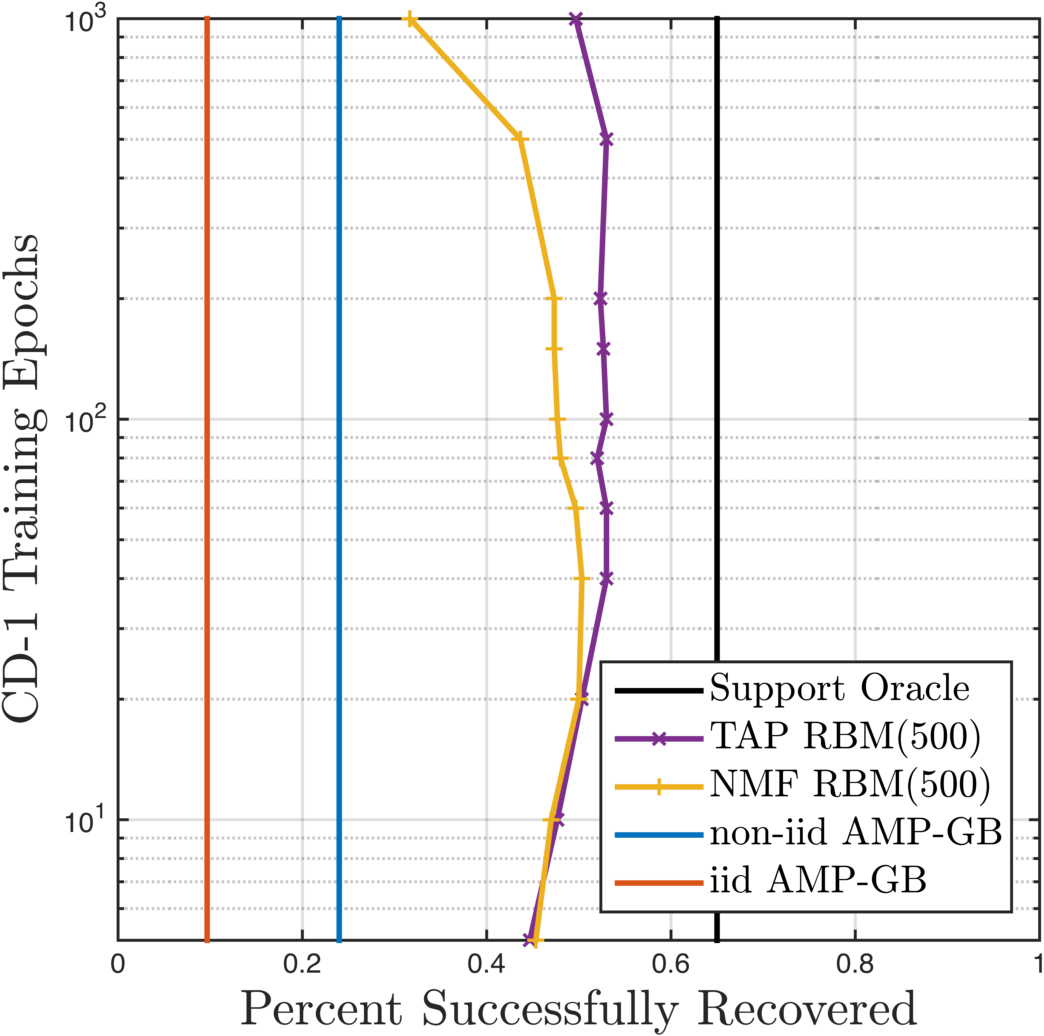}
  \caption{%
    \label{fig:epochCompare}
    \textbf{(Left)}
    Test set reconstruction performance for varying numbers of RBM hidden units.
    One can clearly see in increase in reconstruction performance as the RBM model 
    complexity increases.
    \textbf{(Right)}
    Test set reconstruction performance at $\alpha = 0.2$ as a function of
    the number of epochs, from 5 to 1000, 
    used to learn a 500 hidden unit RBM model. Note the 
    sensitivity of the NMF RBM factorization to model over-fitting.}
  \end{figure}  
In the left panel of Fig. \ref{fig:results}, 
we present the percentage of successfully recovered test set digit
images of the 300 digit images tested. A successful reconstruction is denoted as one
which achieves MSE $\leq 10^{-4}$. It is easily observable from these results that
leveraging a non-\iid support prior does indeed provide drastic performance 
improvements, as even the simple non-\iid version of AMP-GB recovers significantly more
digits than \iid AMP-GB. 
We also see that by using the RBM model
of the support, along with a TAP-based support probability estimation,
we are able to improve upon this simple approach. For example, at
$\alpha = 0.22$, by using the RBM support prior in conjunction with TAP
factorization we are able to recover an additional $56.0\%$ of the test set than 
with no support information, and an additional
$29\%$ than when using only empirical per-coefficient support probabilities.
The fact that both RBM-AMP approaches improve on non-\iid AMP-GB demonstrates
that the learned support correlations are genuinely providing useful information 
during the AMP CS reconstruction procedure. We also note that the 
$2^{{\rm nd}}$-order TAP 
factorization provides reconstruction
performance on this test set 
which is either equal to, or better than, the
NMF factorization, at the cost of an additional matrix-vector multiplication at each iteration of the RBM factorization. To demonstrate how these approaches 
compare with maximum achievable performance, we also show the support oracle 
performance, which corresponds to the percentage of test samples for which 
$\rho \leq \alpha$. The proposed AMP-RBM approach, for the given RBM support 
model, closes the gap to oracle performance. 

In the right panel of Fig. \ref{fig:results}, we show the 
performance of the support estimation in terms of the Matthews 
Correlation Coefficient (MCC) which is calculated from the $2\times 2$ 
confusion matrix of estimated support and the true support. We observe
from this chart that, even though measurement rate might be so low as to
prevent an accurate reconstruction in terms of MSE, the estimated
support of the recovered image may indeed be highly correlated with the true image.
However, the values of the these on-support coefficients will not be 
correctly estimated in this regime. The estimated support may still
be used for certain tasks in this case, such as classification. More visual evidence of 
this effect can be seen in Fig. \ref{fig:vis_results}, were we see, even
to the down to the extreme limit of $\alpha < 0.1$, the support of the 
AMP-RBM recovered images are still quite correlated with the true image.
For non-\iid AMP-GB, the effect of the prior effectively forces the support
to the central region of the image, and for \iid AMP-GB, the support information
is completely lost, having almost zero correlation with the true image.

It is also of interest to analyze the effect of RBM training and model parameters on the performance
of the proposed approach. Indeed, it is curious to note exactly how well trained the RBM must
be in order to obtain the performance demonstrated above. In the right panel of Fig. \ref{fig:epochCompare} we
see that a lightly trained, here on the order of 40 epochs, RBM attains maximal performance, 
showing that an overwrought training procedure is not necessary in order to obtain significant
performance improvements for CS reconstruction. Additionally, in the left panel of Fig. \ref{fig:epochCompare} 
we see that increasing the complexity of the RBM model, and therefore more accurately estimating 
the joint support probability,increases reconstruction performance, showing that more complex RBMs, 
perhaps even stacked RBM models, may have the potential to further improve upon these results.

Lastly, in terms of computational efficiency and scalability, 
the inclusion of the inner RBM factorization loop does increase the computational
burden of the reconstruction in proportion to the number of hidden units and
the number of factorization iterations required for convergence. However, these
iterations are computationally light in comparison to the AMP FPI and so the use
of the RBM support prior is not unduly burdensome.

\section{Conclusion}
	\label{sec:conclusion}
	In this work we show that using an RBM-based prior on the signal support, 
when learned properly, can provide CS reconstruction performance superior 
to that of both simple \iid  and empirical non-\iid sparsity assumptions 
for a message passing approach such as AMP. 
The implications of such an approach are large as these results pave the 
way for the introduction of much more complex and deep-learned priors. Such
priors can be applied to the signal support as we have done here, or 
further modifications can be made to adapt the AMP framework to the use of
RBMs with real-valued visible layers. Such priors would even aid in moving
past the $M=K$ oracle support transition. 

Additionally, a number of interesting generalizations of our approach
are possible.  While the experiments we present here are only concerned
with linear projections observed through an AWGN channel, much more
general, non-linear, observation models can be used moving from AMP to
GAMP \cite{Ran2011}. Our approach can be then readily applied with
essentially no modification to the algorithm.
With the successful application of statistical physics tools
to signal reconstruction, as was done in applying TAP to derive AMP,
similar approaches could be adapted to produce even better learning
algorithms for single and stacked RBMs. Perhaps such future works
might allow for the estimation of the RBM model in parallel with
signal reconstruction.

\section*{Acknowledgment}
We would like thank Andre Manoel for his helpful comments and corrections.
The research leading to these results has received funding from the
European Research Council under the European Union's $7^{th}$
Framework Programme (FP/2007-2013/ERC Grant Agreement 307087-SPARCS).

\clearpage

\bibliographystyle{IEEEtran}
\small{
	\bibliography{references}	
}

\end{document}